\documentclass[twocolumn,longbibliography]{quantumarticle-customized}
\usepackage{float}
\usepackage[pdfpagelabels,pdftex,bookmarks,breaklinks]{hyperref}
\hypersetup{colorlinks,citecolor=blue,urlcolor=blue,linkcolor=blue}

\title{A slightly smaller surface code S gate}
\author{Craig Gidney}
\email{craiggidney@google.com}
\affiliation{Google, Santa Barbara, CA 93117, USA}
\author{Austin Fowler}
\email{agfowler@google.com}
\affiliation{Google, Santa Barbara, CA 93117, USA}

\newcommand{\ket}[1]{\left|#1\right\rangle}

\begin{document}
\maketitle

\begin{abstract}
We cut the volume of surface code $S$ gates by 25\% by omitting a Hadamard gate.
See \autoref{fig:circuit}.
\end{abstract}

\begin{figure}[H]
  \centering
  \includegraphics[width=\linewidth]{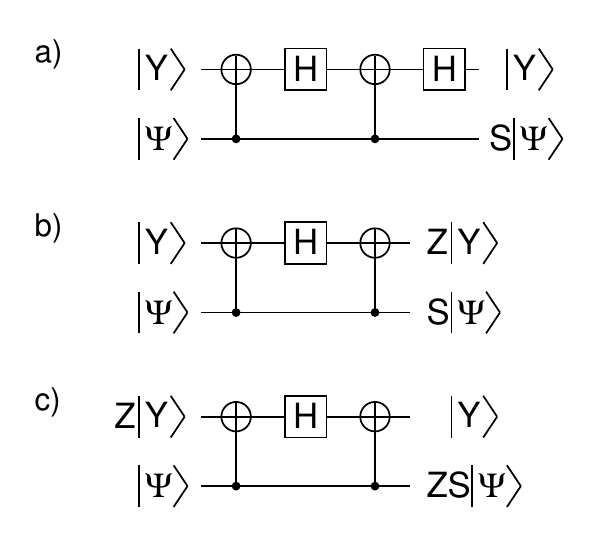}
  \caption{
    Using the ancilla state $|Y\rangle = \frac{1}{\sqrt 2} \big( |0\rangle + i |1\rangle \big)$ to apply an $S$ gate.
    a) Previous $S$ gate circuit \cite{Alif06}, which used a Hadamard gate to correct a phase-flip of the $|Y\rangle$ state.
    b--c) New circuit, which simply allows the phase-flip to happen.
    Unwanted Pauli operators are tracked and corrected in software.
  }
  \label{fig:circuit}
\end{figure}

\begin{figure*}
  \centering
  \makebox[\linewidth]{
    \includegraphics[width=\linewidth]{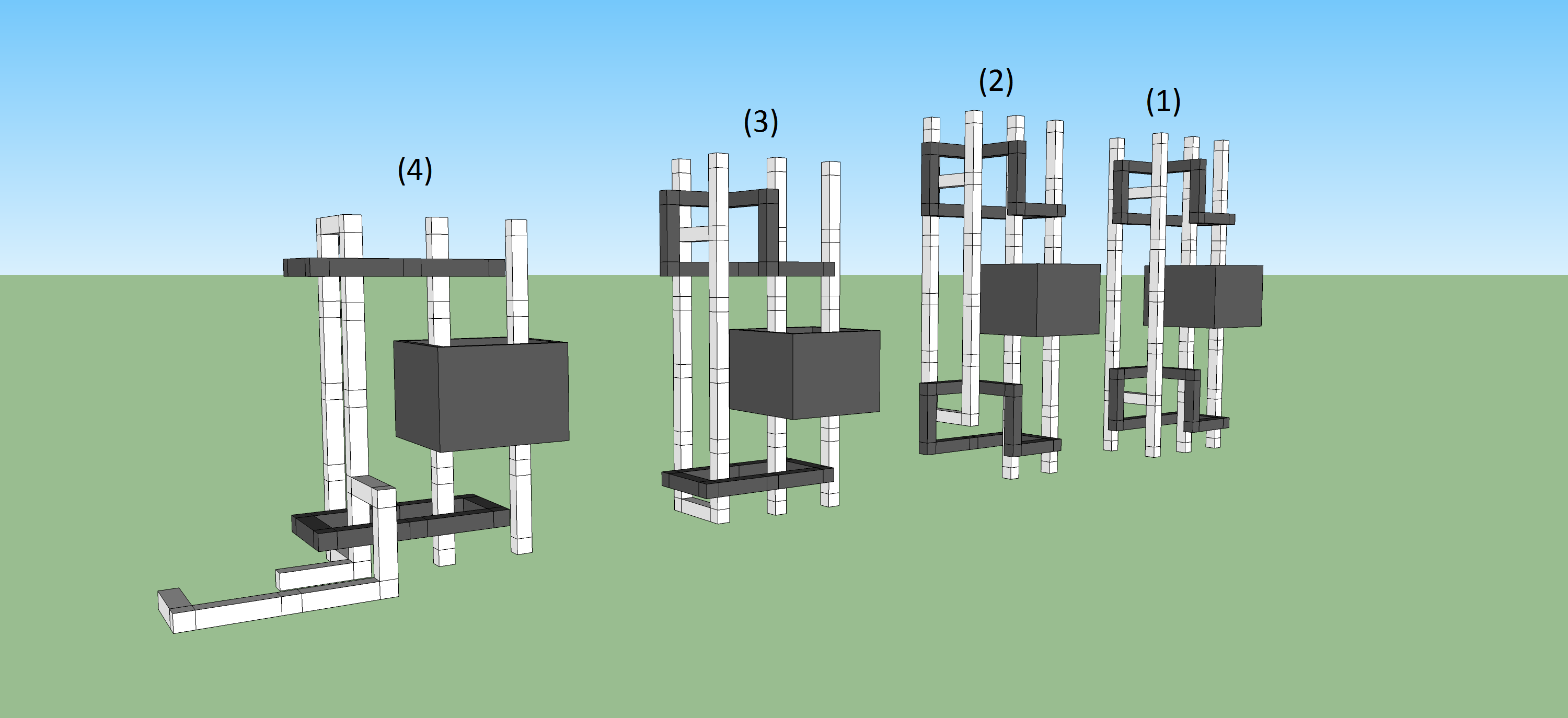}
  }
  \caption{
    Topological quantum error correction protected version of copying a $\ket{Y}$ state.
    1) The structure labelled (1) most closely resembles \autoref{fig:circuit}, with inputs at the bottom and outputs at the top.
    The rightmost two vertical white columns represent the $\ket{Y}$ state, the leftmost two a $\ket{+}$ state.
    The two dark rings are CNOTs, the dark box a Hadamard gate.
    2) Since this is topological quantum error correction, the legs corresponding to $\ket{+}$ state initialization actually achieve nothing and can be removed.
    3) Bottom CNOT has been untwisted and flattened
    4) Outputs have been dragged down to the bottom of the structure to enable repeated copying.
  }
  \label{fig:topology-motion}
\end{figure*}

\begin{figure*}
  \centering
  \makebox[\linewidth]{
    \includegraphics[width=\linewidth]{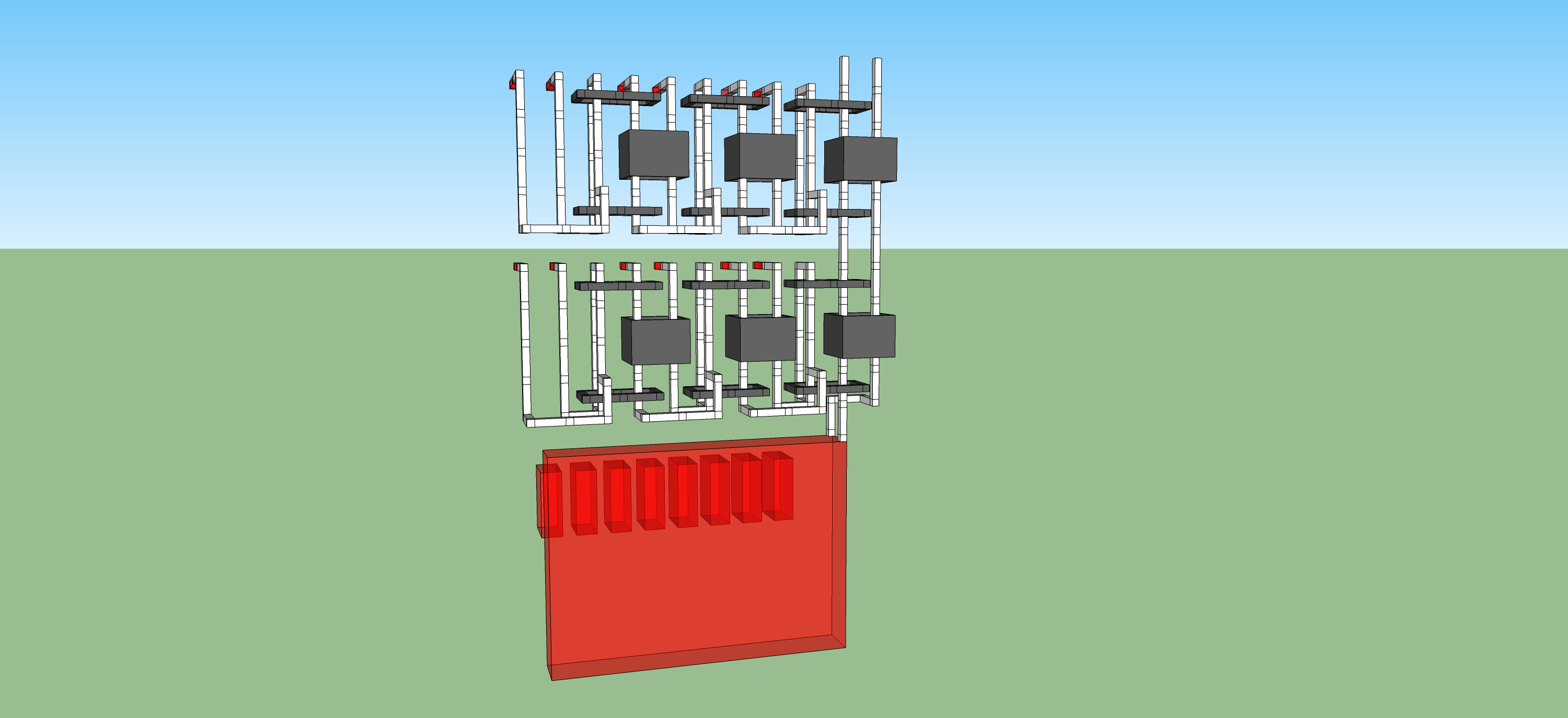}
  }
  \caption{
    The red translucent structure represents two levels of state distillation producing a single $\ket{Y}$ state.
    This state is then copied 6 times, permitting these copies to be consumed during computation, each consumption enabling an $S$ gate.
    This involves nothing more than connecting the rails of a $\ket{Y}$ state to the rails of a data logical qubit.
  }
  \label{fig:distill-use}
\end{figure*}

A wide range of qubit technologies have the potential to implement a 2D array of nearest-neighbor coupled qubits \cite{Bare13,Gamb17,Leik17,Schl11,Laht17}.
Given such a 2D array, the surface code \cite{Brav98,Denn02,Raus07,Raus07d,Fowl12f} has the highest known threshold error rate of approximately 1\%.
Experiments have yet to yield scalable devices with a set of sufficiently low error quantum gates to implement the surface code, and other codes that work in 2D typically have a threshold error rate an order of magnitude lower \cite{Wang09b}.
It is unclear whether gates with sufficiently low error rates to use any code other than the surface code will ever exist.
Therefore, any techniques that make implementing a surface code quantum computer even a little easier deserve be made public.
It is in this spirit that we make this little note available.
See \autoref{fig:circuit}, \autoref{fig:topology-motion}, \autoref{fig:distill-use}.

\bibliographystyle{plain}
\bibliography{citations}

\end{document}